\documentstyle[11pt,aaspp4]{article}
\lefthead{Waxman}
\righthead{GRB after-glow: Short draft}
\begin{document}
\title{GRB after-glow: Supporting the cosmological fireball model,
constraining parameters, and making predictions}
\author{Eli Waxman}
\affil{Institute for Advanced Study, Princeton, NJ 08540;
e-mail: waxman@sns.ias.edu}

\begin{abstract}

Cosmological fireball models of $\gamma$-ray bursts (GRBs) predict delayed 
emission, ``after-glow,'' at longer wavelengths. We present several new results
regarding the model predictions, and show that X-ray to optical observations of
GRB970228 and GRB970402 are naturally explained by the model: The scaling of 
flux with time and frequency agrees with model predictions and requires a power
law distribution of shock accelerated electrons $d\log N_e/d\log\gamma_e=2.3
\pm0.1$ (implying, and consistent with the observed, $t^{-1}$ decline of flux 
observed at a given frequency); The absolute flux value agrees with that 
inferred through the model from observed $\gamma$-ray fluence. The future 
after-glow emission of these bursts is predicted. The observations indicate 
that the ratio of magnetic field to equipartition value and the fraction 
$\xi_e$ of dissipated kinetic energy carried by electrons are not much smaller 
than 1. More frequent observations at a single wavelength, or a wide spectrum 
at a single time, would put strong constraints on these parameters. We show 
that inverse-Compton scattering suppresses X-ray/optical emission at delays 
$t<t_{IC}=10(\xi_e/0.3)^4$hr. Observations therefore imply $\xi_e\le0.3$. The 
strong dependence of $t_{IC}$ on $\xi_e$ implies that $t_{IC}$ may vary widely 
from burst to burst, and that frequent X-ray/optical observations at 
$t\sim t_{IC}$ would determine $\xi_e$. For $\xi_e\sim0.2$, inverse-Compton 
emission dominates during the first 2hr, producing $>1$GeV photons and 
providing a natural explanation to the delayed GeV emission observed in 
several strong bursts.

\end{abstract}

\section{Introduction}

Recent observations of $\gamma$-ray bursts (GRBs)
suggest that they originate from cosmological sources (\cite{cos1,cos2}). 
General phenomenological considerations indicate that the 
bursts are produced by the dissipation of the kinetic energy of a 
relativistic expanding fireball (see \cite{scen1}, \cite{scen2} for reviews). 
After producing the main GRB, the expanding cooling fireball is expected to
produce delayed emission at longer wavelengths (\cite{ag1,ag2}). 
Such optical and X-ray after-glow has possibly been detected for
GRB970228 and GRB970402: Variable X-ray sources have been
detected within the $3\arcmin$ GRB error boxes, 
fading on a time scale of days (\cite{bepg},b, \cite{bepg1,bepx1}); 
A variable optical source, fading on a similar time scale, 
was detected within the $50\arcsec$ error box of the
X-ray source associated with GRB970228 (\cite{opt1,opt2}). 

The GRB fireball model depends on several parameters: $E$, the total fireball
energy; $\eta^{-1}$, the fraction of energy initially carried by (proton) rest
mass; $T$, the energy emission duration; $\xi_e$, the fraction of dissipated
kinetic energy carried by electrons; $\xi_B$, the ratio of magnetic field to 
equipartition field; $p$, the spectral index of shock accelerated electrons
($dN_e/d\gamma_e\propto\gamma_e^{-p}$).
In sec. 2 below we derive the after-glow predictions of the fireball model.
Our results follow closely those of M\'esz\'aros \& Rees (1997), the main new 
results being: 
(i) After-glow due to interaction with the inter-stellar medium (ISM) is
determined by $E$, $\xi_e$, $\xi_B$, and $p$ only (in particular, it is 
independent of $\eta$ and of whether the energy is emitted impulsively or 
as a wind): The flux peaks at a frequency $\nu_m=C_1E^{1/2}\xi_e^2\xi_B
t^{-3/2}$, and for $\nu\ge\nu_m$ $F_\nu=C_2E\xi_Bn^{1/2}(\nu/\nu_m)^{-\alpha}$,
where $\alpha=(p-1)/2$ and $n$ is the ISM density;
(ii) Although $p$ is expected to be similar for the GRB and the after-glow,
$\alpha$ during the after-glow is expected to be smaller by $1/2$ due to 
increase in cooling time. For the GRB $\alpha\sim1$, implying $\alpha\sim0.5$
for the after-glow. This also implies a decline in flux 
at a fixed wavelength much slower than the $t^{-3/2}$ decline 
obtained assuming similar $\alpha$ for the GRB and after-glow; 
(iii) Although the delayed X-ray/
optical emission must result from synchrotron emission of accelerated ISM
electrons, inverse-Compton emission may dominate during
the first hours--days, suppressing the X-ray/optical emission.

Observations are compared with model predictions in sec. 3. 
In sec. 4 we summarize the comparison results, discuss the observational 
constraints on the model and point out future observations that may further 
constrain model parameters, give the model predictions for
future GRB970228 and GRB970402 after-glow, and
discuss the sensitivity of the results to underlying assumptions.

\section{Model and Parameters}

Whatever the sources of GRBs are, observations suggest the following scenario
for the emission of the observed $\gamma$-rays. A compact, $r_0\sim10^7$cm,
source releases an energy $E$ comparable to that observed in $\gamma$-rays,
$E\sim10^{51}$erg, over a time $T<100$s. The large energy density in the 
source results in an optically thick plasma that expands and accelerates to
relativistic velocity. After an initial acceleration phase, the fireball
energy is converted to proton kinetic energy. A cold shell of 
thickness $cT$ is formed and continues to expand with time independent
Lorentz factor 
$\gamma=\eta\sim300$. The GRB is produced once the kinetic energy
is dissipated at large radius, $r>10^{13}$cm, and radiated as $\gamma$-rays
through synchrotron and possibly inverse-Compton emission of shock accelerated 
electrons. 

Dissipation may occur due to internal collisions within the shell 
(\cite{int1,int2}), or due to 
collision with the inter-stellar medium (ISM) (\cite{coll}). 
ISM collision 
is expected in both cases, since following internal collisions, which
convert part of the energy to radiation and result
from variations in $\gamma$ across the expanding shell, 
the shell rapidly cools and
continues to expand with approximately uniform Lorentz factor $\gamma=\eta$.
At first, the ISM influence
on the shell is negligible. As the cold shell propagates with $\gamma=
\eta$, it drives a shock into the ISM. This shock propagates with $\gamma_s=
2^{1/2}\gamma$, and behind it the (rest frame) number and energy densities
are $n'=4\gamma n$ and $e'=4\gamma^2nm_pc^2$ respectively, 
where $n$ is the ISM number
density ahead of the shock. The width of the shock heated ISM shell
is $r/4\gamma^2$, where $r$ is the shell radius.
It is usually assumed that the shell is influenced by the 
ISM once the heated ISM energy,
$4\pi r^2(r/4\gamma^2)\gamma^2e'$, is comparable
to $E$. This occurs at a radius
$r_E=(E/4\pi\eta^2nm_pc^2)^{1/3}$.
However, Sari \& Piran (1995) pointed out that 
the ISM may influence the shell at an earlier time,
as it may drive a relativistic shock wave into the shell at $r<r_E$. When 
such a shock forms the shell is no longer ``cold''.
A relativistic backward shock is formed once
the energy density $e'$ implies a relativistic energy density in the shell,
$e'/n'_sm_pc^2\sim1$, where $n'_s$ is the cold shell number density.
This occurs at 
\begin{equation}
r_R=2\times10^{15}\left({E_{51}\over T_{10}\eta_{300}^4n_1}\right)^{1/2}{
\rm cm},
\end{equation}
where $E=10^{51}E_{51}$erg, $\eta=300\eta_{300}$, $n=n_1{\rm cm}^{-3}$,
$T=10T_{10}$s. If $r_R>r_E$, a relativistic backward shock is not formed,
and the shell loses its energy to the heated ISM at $r_E$. If $r_R<r_E$,
most of the shell kinetic energy is converted to thermal energy once the
backward shock crosses the shell. 
This occurs at (\cite{SP})
\begin{equation}
r_c=\left({ET\over4\pi nm_pc}\right)^{1/4}=
10^{16}\left({E_{51}T_{10}\over n_1}\right)^{1/4}{\rm cm}.
\label{rc}
\end{equation}
At this time the heated shell and ISM move with Lorentz factor
\begin{equation}
\gamma_c=\left({r_c\over2Tc}\right)^{1/2}=
140\left({E_{51}\over T_{10}^3 n_1}\right)^{1/8}.
\label{gc}
\end{equation}

The ratio of $r_R$ to $r_E$ is $r_R/r_E=
0.25E_{51}^{1/6}/T_{10}^{1/2}n_1^{1/6}\eta_{300}^{4/3}$.
Since $\eta\ge300$ is required in order for the pair production optical depth 
to be sufficiently small (e.g. \cite{eta}), $r_R<r_E$
for $T>1s$. In this case, most of the kinetic energy is converted at $r=r_c$ to
thermal energy of the heated ISM and shell (it can be shown that at this point
the shell thickness is similar to that of the heated ISM region, 
$r_c/4\gamma_c^2$, and therefore the ISM and heated shell carry 
similar energy). Since the conversion
of kinetic energy to thermal energy occurs (as seen by the
observer) 
over a time scale of $r_c/2\gamma_c^2=T$, the duration of the GRB is
$T$ regardless of whether it
is produced by internal collisions or by ISM collision.
If $T<1$s and $r_R>r_E$, only the dissipation at $r_E$ produces the GRB. 
In this case, the burst duration is $r_E/\eta^2c$. However, it is straight
forward to show 
that if we require the duration of the burst to be $T_{du}$, then $r_E$
and the shell Lorentz factor $\eta$ are given by the above equations
for $r_c$ and $\gamma_c$ with $T$ replaced with $T_{du}$. Thus, the conditions
at the ISM heated shock at the end of the main burst, which determine
the following after-glow, are
always given by (\ref{rc}) and (\ref{gc}), with $T$ interpreted as the main
GRB duration (the duration over which most of the $\gamma$-ray energy is
observed). Note that the dependence of $r_c$ and $\gamma_c$ on parameters
is very weak, and that they are mainly determined by $E$ (and the observable
$T$).

At $r>r_c$, the shell decelerates and a relativistic shock continues to 
propagate into the ISM. The electrons accelerated during the initial
collision at $r=r_c$ rapidly lose their energy (\cite{ag2}). 
However the shock driven into the ISM continuously produces relativistic 
electrons that may produce the delayed radiation observed on time scales of
days to months.
As before, the shell expanding with Lorentz factor $\gamma$ drives
a shock with 
$\gamma_s=2^{1/2}\gamma$ into the ISM.  
The heated ISM energy is $4\pi r^2(r/4\gamma^2)
\gamma^2e'=E/2$. This determines the evolution of $\gamma$,
\begin{equation}
\gamma=\gamma_c(r/r_c)^{-3/2}.
\label{gt}
\end{equation}
Light emitted from the shell at radius $r$ is observed over duration
$t=r/2\gamma^2c=T(r/r_c)^4$.
This completely determines the evolution of the shell of baryons. In order
to calculate the synchrotron emission from the heated ISM shell we need
to determine the magnetic field and electron energy. We assume that the
magnetic field (in the shell rest frame) is a fraction $\xi_B$ of the
equipartition value, $B'=\xi_B(8\pi e')^{1/2}$, and that the electrons
carry a fraction $\xi_e$ of the energy. Since the Lorentz factor associated
with the thermal motion of protons in the shell rest frame is $\gamma$,
this implies that the Lorentz factor of the random motion of a typical
electron in the shell rest frame is $\gamma_{em}=\xi_e\gamma m_p/m_e$. 
With these
assumptions, and using eqs. (\ref{rc}), (\ref{gc}) and (\ref{gt}), 
we
find that the observed frequency of synchrotron emission from typical
electrons, $\nu_m\simeq\gamma\gamma_{em}^2eB'/m_ec$, varies with time as
\begin{equation}
\nu_m=10^{16}\xi_e^2\xi_BE_{51}^{1/2}t_{\rm day}^{-3/2}{\rm Hz},
\label{nu}
\end{equation}
where $t=1t_{\rm day}$days. 
This is the frequency where synchrotron emission peaks.
Emission at higher frequencies is produced by electrons with energy higher 
than typical. The ratio of synchrotron cooling time, $t_s=6\pi m_ec/
\sigma_T\gamma_eB'^2$, to rest frame expansion time, $t_d=r/\gamma c$, is
\begin{equation}
t_s/t_d=0.03\xi_e^{-1}\xi_B^{-2}E_{51}^{-1/2}t_{\rm day}^{1/2}n_1^{-1/2}.
\label{tstd}
\end{equation}

From eq. (\ref{tstd}), we expect the synchrotron cooling time to be larger
than the dynamical time for long delay after-glow. As we show in sec. 3, for
GRB970228 the parameters are such that for most of the after-glow
$t_s>t_d$. In this case, the flux at $\nu_m$ 
can be obtained in the
following way. The number of radiating electrons at a given time
is $4\pi r^3n/3$, each producing synchrotron photons at a rate (in the
shell rest frame) $P_S=\gamma_{em}m_ec^2/t_s(h\nu_m/\gamma)$ (the frequency
at the rest frame is smaller by a factor $\gamma$).
Since $\nu_m\propto r^{-6}$,
emission over a range $\Delta\nu_m$ occurs over
radius range $\Delta r=\Delta\nu_mr/6\nu_m$, which corresponds to a
time in the rest frame $\Delta t=\Delta r/\gamma c$. Thus, the energy
(in the observer frame) emitted over the frequency range $\Delta\nu_m$ is
$\Delta E=(2\pi/9)r^4n(\gamma_{em}m_ec^2/t_sc)\Delta\nu_m/\nu_m$. This energy
is observed over a time $r/2\gamma^2c$. Thus, the observed intensity at the 
frequency $\nu_m$ is
\begin{equation}
F_{\nu_m}={\Delta E\over\Delta\nu_mr/2\gamma^2c}=
{2\over27\sqrt{2\pi}}{\sigma_Tm_ec\over m_p^{1/2}e}\sqrt{n}
{\xi_BE\over4\pi d^2},
\label{Fm}
\end{equation}
where $d$ is the source distance.
If the electron distribution has a high energy tail, $dN_e/d\gamma_e\propto
\gamma_e^{-p}$ for $\gamma_e>\gamma_{em}$, then for $\nu>\nu_m$ (defining 
$\alpha\equiv(p-1)/2$):
\begin{equation}
F_\nu=F_{\nu_m}[\nu/\nu_m(t)]^{-\alpha}.
\label{Ft}
\end{equation}

Note that eq. (\ref{tstd}) implies that at early times $t_s<t_d$ (unless
the efficiency of converting kinetic energy to radiation is very low).
We show in sec. 3 that for
GRB970228 the parameters are such that indeed $t_s\ll t_d$ at early times.
In fact, it has been shown by \cite{SPN} that $t_s\ll t_d$ is typically
required for the GRB emission. 
Thus, although $p$ is likely to be similar for the
GRB and for the after-glow, since in both cases high energy electrons
are assumed to be produced by shock acceleration, the spectral index $\alpha$
for the GRB is expected to be larger by $1/2$.

Let us now consider inverse-Compton emission from the accelerated electrons.
The ratio of inverse-Compton to synchrotron emission, $P_{IC}/P_S$, is
given by $P_{IC}/P_S=U_S/U_B$, where $U_S$ and $U_B$ are the synchrotron
photons and magnetic field energy density. For $t_d<t_s$, and when 
emission is dominated by the synchrotron process, the total energy
in synchrotron photons is a fraction $t_d/t_s$ of the electron energy, and
the energy density $U_S$ is a fraction $t_d/4t_s$ of the electron energy 
density (the factor of 4 is due to the fact that the rest frame
thickness of the shell is $r/4\gamma$, while the radiation  
occupies a shell of thickness $r/\gamma$). Using (\ref{tstd}) we then
obtain
\begin{equation}
P_{IC}/P_S=9\xi_e^2n_1^{1/2}E_{51}^{1/2}t_{\rm day}^{-1/2}.
\label{ic}
\end{equation}
The inverse-Compton frequency is a factor $\gamma_e^2$ larger than the
synchrotron frequency,
\begin{equation}
\nu_{IC}=7\times10^{23}\xi_e^4\xi_B
n_1^{-1/4}E_{51}^{3/4}t_{\rm day}^{-9/4}{\rm Hz}.
\label{icnu}
\end{equation}
Eq. (\ref{ic}) implies that $\xi_e>0.3$ is required in order for inverse-
Compton emission to be important at $\sim1$day. For $\xi=0.3$, however,
(\ref{icnu}) implies that in order to obtain emission at optical wavelength
$\xi_B<10^{-6}$ is required. It is easy to see from eq. (\ref{Fm}) that this
would require $E$ to be many orders of magnitude larger than the observed
$\gamma$-ray energy in order for an optical flux to be detectable (see
also sec. 3 below). Since the observed $\gamma$-ray 
energy is already challenging for most sources, this is not likely. 
Although inverse-Compton emission can 
not be important when optical emission is seen, it will dominate
at early times (unless $\xi_e\ll1$), shifting the predicted synchrotron 
after-glow emission to energy well above X-ray/optical. This is discussed
in more detail below.

\section{GRB970228 and GRB970402}

Let us first discuss GRB970228. Eq. (\ref{Ft}) shows, that
the time and frequency dependence of $F_\nu$ is determined by
$p$. For GRB970228, X-ray flux of $(2.8\pm0.4)\times10^{-28}{\rm erg/s\,cm}
^2$ was observed 10hr after the burst at the 2-10keV band (\cite{bepg}), 
corresponding
to $F_\nu=(2\pm0.3)\times10^{-30}{\rm erg/s\ Hz\ cm}^2$ at $\nu=10^{18}$Hz,
and a flux $F_\nu=1.2\times10^{-28}{\rm 
erg/s\ Hz\ cm}^2$ at $\nu=3.4\times10^{14}$Hz ($m_I=20.6$)
was detected after 21hr (\cite{opt1}). Using these values in (\ref{Ft})
we obtain $\alpha=0.63$, or $p=2.3$. Since the frequency and flux range are
large, $\alpha$ is determined with good accuracy: The accuracy of the optical
fluxes is not quoted; Assuming an accuracy of $20\%$, the ($1\sigma$) error
in $\alpha$ is $\sim0.05$. At 21hr delay, there is a 
simultaneous detection $m_V=21.3$, corresponding to 
$F_\nu=1.1\times10^{-28}{\rm erg/s\ Hz\ cm}^2$ at $\nu=5.5\times10^{14}$Hz.
This is consistent with $\alpha=0.63$. However, since the fluxes are 
probably accurate to only $\sim10\%$, and since the frequency difference is 
small, this is not a strong test for the model.

More stringent tests are obtained by observations made at larger delays.
With $\alpha=0.63$, the flux at a given frequency should drop with time
as $t^{-1}$. Following the initial detections, observations in both X-ray 
(2--10keV) and optical (V, I) bands were carried out at a time delay larger
by a factor of 10 compared to the detection delay (87hr for X-ray, 9days
for optical). Although
no flux is quoted for the second X-ray observation, it is mentioned that the
flux dropped by a factor of nearly 20. Optical observations yielded
limits of $m_V>23.6$
and $m_I>22.2$ at 9day, implying a drop in flux by a factor larger than
8 an 4 respectively. These observations are consistent with the model. 
HST detected the source in the optical band at 38day delay with $m_V=26.0\pm
0.3$ and $m_I=24.6\pm0.3$ (\cite{opt2}). Using the fluxes measured at 21hr
delay, and the $t^{-1}$ scaling, the fluxes predicted by the model at 38day
are $m_V=25.4$ and $m_I=24.7$, 
in remarkable agreement with the observed fluxes
(Note that a $t^{-3/2}$ decline would predict $m_V=27.4$, $m_I=26.7$).

The $\gamma$-ray fluence of the main GRB 
is $2\times10^{-6}{\rm erg/cm}^2$, emitted over $\sim4$s duration with a 
spectral peak at $100-150$keV (\cite{gamma}) [BeppoSax detected 2 additional
sub bursts and some lower flux level extending over 80s (\cite{bepg}); 
however, the peak fluxes of subsequent bursts are lower
by a factor of 10, and the total emission during the 80s is at most twice that
of the main burst (\cite{gamma})]. 
This fluence determines the value of $f_\gamma E/4\pi d^2$,
where $f_\gamma$ is the fraction of energy converted to $\gamma$-rays
($E=10^{51}$erg for $f_\gamma=1$ and $d=2$Gpc, i.e. $z\simeq0.5$). Using this
value in (\ref{Fm}) we have
\begin{equation}
F_{\nu_m}^\gamma=1.7\times10^{-27}\sqrt{n}f_\gamma^{-1}\xi_B
{\rm erg/s\ Hz\ cm}^2.
\label{Fg}
\end{equation}
The value of $F_{\nu_m}$ from X-ray and optical detections can be
derived from (\ref{Ft}) if $\nu_m(t)$ is known for some 
$t$. This 
can be determined by frequent observations of at a fixed frequency or
a by obtaining a wide spectrum at fixed time. 
Such observations are not available
for GRB970228. However,
we can place limits. The optical detections imply that the time $t_I$ at
which the synchrotron peak is at $\nu_m=3.4\times10^{14}$ is $t_I<21$hr.
Using this in (\ref{Fm}) we have
\begin{equation}
F_{\nu_m}^{I}=1.2\times10^{-28}t_{I,21}^{-1}{\rm erg/s\ Hz\ cm}^2.
\label{FI}
\end{equation}
where $t_I=21t_{I,21}$hr.
This is in remarkable agreement with the $\gamma$-ray inferred flux.

Comparing (\ref{Fg}) and (\ref{FI}) and using (\ref{nu}) we find
\begin{equation}
\xi_B=0.1t_{I,21}^{-1}f_\gamma n_1^{-1/2},\quad
\xi_e=0.6f_\gamma^{-1/4}n^{1/4}_1d_2^{-1/2}t^{5/4}_{I,21},
\label{xiB}
\end{equation}
where $d=2d_2$Gpc.
Thus, the magnetic field should be close to equipartition, unless the 
conversion of energy to $\gamma$-rays is very inefficient.
The value of $\xi_e$ depends mainly on $t_I$. Both $\xi_e$ and $\xi_B$
would be strongly constrained by observationally determining $t_I$.
Using (\ref{xiB}) in (\ref{tstd}) we find that $t_s>t_d$
for $t>10^3f_\gamma^3 n^{-1/2}E_{51}^{1/2}t_{I,21}^{-3/2}$s. This justifies 
our assumption that for most
of the after-glow we can assume $t_s>t_d$, and that $t_s<t_d$ at earlier
times, with the implications we mentioned for $\alpha$ and $p$.

An argument similar to that following (\ref{icnu}) for the X-ray detection 
at 10hr, shows that inverse-Compton emission can not be important at this
time. Using (\ref{ic}) this implies $\xi_e\le0.3E_{51}^{-1/4} n_1^{-1/4}$.
For a given value of $\xi_e$, the time up to which inverse-
Compton dominates the emission is
\begin{equation}
t_{IC}=10n_1E_{51}(\xi_e/0.3)^4\,{\rm hr}.
\label{tIC}
\end{equation}
[Eq. (\ref{tIC}) is valid for $\xi_e>0.15E_{51}^{-2/13}
n_1^{-3/13}f_\gamma^{6/13}$, since we have implicitly
assumed that $t_s>t_d$ for $t\ge t_{IC}$ in our calculation, which does
not hold for lower $\xi_e$ values.]
For $t<t_{IC}$, emission is dominated by the inverse-Compton process, which
produces very high energy photons and suppresses the 
X-ray/optical after-glow. 

For GRB970402 only X-ray after-glow has (so far) been reported (\cite{bepx1}). 
It is 
therefore impossible to constrain the relevant parameters for this burst. 
However, it is remarkable that the detected after-glow is consistent with
the model described above. The $\gamma$-ray peak flux of GRB970402 is
0.08 times that of GRB970228. Assuming both bursts have similar intrinsic 
properties ($E,\,p,\,\xi_e$ and $\xi_B$), this implies that the X-ray flux
of this burst measured at 11hr delay (the flux was integrated over 8 to 15hr,
and for $t^{-1}$ decline the average flux corresponds to the flux at 11hr)
should be 0.07 times 
the GRB970228 flux at 10hr, i.e. $2\times10^{-13}{\rm erg/cm}^2{\rm s}$, 
in agreement with the measured flux, $1.5\pm0.5\times10^{-13}
{\rm erg/cm}^2{\rm s}$. The non-detection of X-ray emission at 32hr delay
implies that the flux decreased by more than a factor of 3 compared to the
11hr delay, in agreement with the model prediction.

\section{Discussion}

We have shown that GRB after-glow observations are naturally explained by the
fireball model: (i) The functional dependence of the flux on time and frequency
is determined only by the spectral index $p$ of shock accelerated 
electrons. GRB970228 after-glow observations are consistent with the
predicted scaling (\ref{Ft}), and require $p=2.3\pm0.1$, well within the range 
expected for shock acceleration; (ii) The absolute value of the observe
flux (\ref{FI}) is consistent with that inferred through the model from 
the $\gamma$-ray fluence (\ref{Fg}). Apart from determining $p$, the observed
value of the flux and the observed normalization of the $\nu_m\propto t^{-3/2}$
relation provide, through eqs. (\ref{nu}) and (\ref{Fm}), two constraints
on model parameters  $\{\xi_e,\xi_B,E\}$. These constraints are given for 
GRB970228 in eq. (\ref{xiB}), where the fireball energy
$E$ is replaced with $f_\gamma$, 
the fraction
of energy converted to the observed $\gamma$-ray fluence, $n$ is the ISM 
density, and $d$ is the burst distance. 
The parameter $t_I$ that appears in (\ref{xiB}) is the time at which $\nu_m$
is in the I band, which can only be constrained to $t_I\le21$hr for this burst.
This parameter may be better constrained by frequent optical observations at 
shorter delays, or by obtaining a wide spectrum at a single time. 
This would allow to put strong constraints on the least
well known model parameters, $\xi_e$ and $\xi_B$. Current observations indicate
that both can not be much smaller than unity. Note that although the
dependence of the $\xi_e$ constraint on $d$ is weak,
local GRBs with $d\sim100$kpc instead of $d\sim1$Gpc would imply $\xi_e\gg1$,
which is of course impossible. Thus, the observed after-glow can not be
explained by the model considered here for local GRBs. 

Using (\ref{Ft}), the derived value of $\alpha$ and 21hr V band observation,
the future behavior of GRB970228 should follow
\begin{equation}
m=m_0+2.5\log(t/t_0),
\label{mag}
\end{equation}
where $m_0=25.1,\,24.2,\,23.7,\,22.3$ in V, I, J, K bands 
respectively with $t_0=30$day ($t$ measured from the GRB). For GRB970402
there is only one X-ray measurement, and it is impossible to determine whether
the parameters $\{\xi_e,\xi_B,p\}$ are the same as for GRB970228. If, however, 
these parameters do not vary much between bursts, (\ref{mag}) should also
describe GRB970402 with $m_0=24.2,\,23.2,\,22.8,\,21.3$ in V, I, J, K bands 
with $t_0=1$day.

For the parameters required for GRB970228
the emission is dominated by the inverse-Compton process, which
produces very high energy photons and suppresses the 
X-ray/optical after-glow, for $t<t_{IC}=10(\xi_e/0.3)^4$hr (for
$\xi_e\gtrsim0.1$). The observations
therefore imply 
$\xi_e\le0.3$. The strong dependence of $t_{IC}$ on $\xi_e$ implies
that $t_{IC}$
may vary widely from burst to burst. 
On the other-hand, measuring $t_{IC}$ by observing the appearance (absence) of
X-ray/optical after-glow after (before) $t_{IC}$
would provide a strong constraint on $\xi_e$.
For $\xi_e\sim0.2$, the inverse-Compton scattered
photons have a characteristic energy of $\sim1$GeV during the first $\sim2$hr
after the burst, therefore providing a natural explanation for the
delayed GeV emission observed in several strong bursts (\cite{GeV}).
X-ray and optical after-glow observations therefore seem to support the 
suggestion of \cite{dG} that the delayed GeV emission is due
to inverse-Compton scattering of ISM electrons heated by the shell collision
with the ISM. Note,
however, that in \cite{dG} the frequency and duration of inverse-Compton
emission is strongly dependent not only on $\xi_e$ but also 
on the initial fraction of fireball
energy carried by rest mass, $\nu_{IC}\propto\eta^6$ and
$t_{IC}\propto\eta^4$. As we have shown here, the after-glow is
independent of $\eta$ [cf. eq. (\ref{nu})--(\ref{icnu})].

An underlying implicit assumption of the analysis presented in sec. 2, is 
that $\xi_e$, $\xi_B$ and $p$ are constant during the shell deceleration 
period. Since the ISM shock is relativistic up to a month
after the GRB event, $\gamma_s=2(E_{51}/n_1)^{1/8}(t/30{\rm day})^{-3/8}$ 
[cf. eq. (\ref{gc}), (\ref{rc}), (\ref{gt})], 
it is not unreasonable that the parameters $p$, 
$\xi_e$ and $\xi_B$, which are determined by the physical processes at the 
shock, would remain constant during this time. 
If these parameters are time dependent, the predicted scaling
(\ref{Ft}) would not hold. The good agreement of GRB970228
after-glow observations with (\ref{Ft}) indicate that this is not the case. 
However, it should be noted that at delays much larger than months, the
shell is no longer relativistic and the predicted scaling laws may no longer
hold. Another implicit assumption that was made is 
that the expanding shell is spherical. 
The results are valid, however, also for a jet of opening angle $\theta$, 
as long as $\theta>1/\gamma$. At times for which $\gamma<1/\theta$ the flux
would decrease much faster than predicted for a spherical shell. Once again,
GRB970228 after-glow observations imply that opening angle is not small.
It should be emphasized, that since (\ref{xiB}) imply that both $\xi_e$
and $\xi_B$ are not much smaller than 1, a significant 
decrease with time in these 
parameters, or similarly a decrease in $\gamma$ below $1/\theta$, would
both result in fluxes much lower than observed. 

\acknowledgements
I thank J. N. Bahcall and P. Kumar for helpful comments. This research
was partially supported by a W. M. Keck Foundation grant 
and NSF grant PHY 95-13835.

{}

\end{document}